\documentclass[
 reprint,
superscriptaddress,
amsmath,amssymb,amsfonts,
aps,
prl,
]{revtex4-1}

\usepackage{graphicx}
\usepackage{dcolumn}
\usepackage{bm}
\usepackage{bbm}
\usepackage[utf8]{inputenc}
\usepackage{amsthm}
\usepackage{mathrsfs}
\usepackage{tikz}
\usepackage{physics}
\usepackage{comment}
\usepackage{xcolor}
\usepackage{blindtext}
\usepackage{pgfplots,pgfplotstable}
\usepgfplotslibrary{polar}
\usepgfplotslibrary{colormaps}
\pgfplotsset{compat=newest}
\usepackage{hyperref}

\definecolor{DTgreen}{rgb}{.10, .4, .3}

\definecolor{HS}{rgb}{0.0,0.5,0.0}
\definecolor{SN}{rgb}{1, 0.5, 0.31}
\definecolor{SSN}{rgb}{0.87, 1.0, 0.0}
\definecolor{OSN}{rgb}{1.0, 0.44, 0.37}

\providecommand{\abs}[1]{\left\lvert#1\right\rvert}

\newcommand{\e}{\mathrm{e}}
\newcommand{\ii}{\mathrm{i}}

\newcommand{\Vi}{V_{\mathrm{in}}}
\newcommand{\Vo}{V_{\mathrm{out}}}
\newcommand{\hVi}{\hat{V}_{\mathrm{in}}}
\newcommand{\hVo}{\hat{V}_{\mathrm{out}}}

\theoremstyle{theorem}

\theoremstyle{definition}

\begin{document}

\title{Heisenberg scaling precision in multi-mode distributed quantum metrology}

\author{Giovanni Gramegna}
\email{giovanni.gramegna@ba.infn.it}
\affiliation{Dipartimento di Fisica and MECENAS, Università di Bari, I-70126 Bari, Italy }
\affiliation{INFN, Sezione di Bari, I-70126 Bari, Italy}

\author{Danilo Triggiani}
\email{danilo.triggiani@port.ac.uk}
\affiliation{School of Mathematics and Physics, University of Portsmouth, Portsmouth PO1 3QL, UK}
\author{Paolo Facchi}
\affiliation{Dipartimento di Fisica and MECENAS, Università di Bari, I-70126 Bari, Italy }
\affiliation{INFN, Sezione di Bari, I-70126 Bari, Italy}
\author{Frank A. Narducci}
\affiliation{Department of Physics, Naval Postgraduate School, Monterey, CA, United States 
}
\author{Vincenzo Tamma}
\email{vincenzo.tamma@port.ac.uk}
\affiliation{School of Mathematics and Physics, University of Portsmouth, Portsmouth PO1 3QL, UK}
\affiliation{Institute of Cosmology and Gravitation, University of Portsmouth, Portsmouth PO1 3FX, UK}

\date{\today}

\begin{abstract}
We propose an $N$-photon Gaussian measurement scheme which allows the estimation of a parameter $\varphi$ encoded into a multi-port interferometer with a Heisenberg scaling precision (i.e.\ of order $1/N$). In this protocol, no restrictions on the structure of the interferometer are imposed other than linearity and passivity, allowing the parameter $\varphi$ to be distributed over several components.
In all previous proposals Heisenberg scaling has been obtained provided that both the input state and the measurement at the output
are suitably adapted to the unknown parameter $\varphi$. This is a serious drawback which would require in practice the use of iterative procedures with a sequence of trial input states and measurements, which involve an unquantified use of additional resources.
Remarkably, we find that only one stage has to be adapted, which leaves the choice of the other stage completely arbitrary. We also show that our scheme is robust against imperfections in the optimized stage. Moreover, we show that the adaptive procedure only requires a preliminary classical knowledge (i.e to a precision $1/\sqrt{N}$) on the parameter, and no further additional resources. As a consequence, the same adapted stage can be employed to monitor with Heisenberg-limited precision any variation of the parameter of the order of $1/\sqrt{N}$ without any further adaptation.

\end{abstract}

\pacs{Valid PACS appear here}

\maketitle

\paragraph{Introduction} Due to the discreteness of all natural phenomena, 
the error in the estimation of a physical parameter $\varphi$ through a measurement employing $N$ probes (e.g. photons, electrons) is strongly limited by the so-called ``shot noise'' factor of $1/\sqrt{N}$. However, it has been proven that quantum features such as  entanglement and squeezing can be exploited to go beyond the shot-noise limit and reach a precision of order $1/N$, which is the so-called Heisenberg limit~\cite{Giovannetti2004,Giovannetti2006,Dowling2008,Giovannetti2011,Dowling2015,Shapiro84,Wineland92,Maccone2019}. 

\begin{figure}[t]
	\centering
	\includegraphics{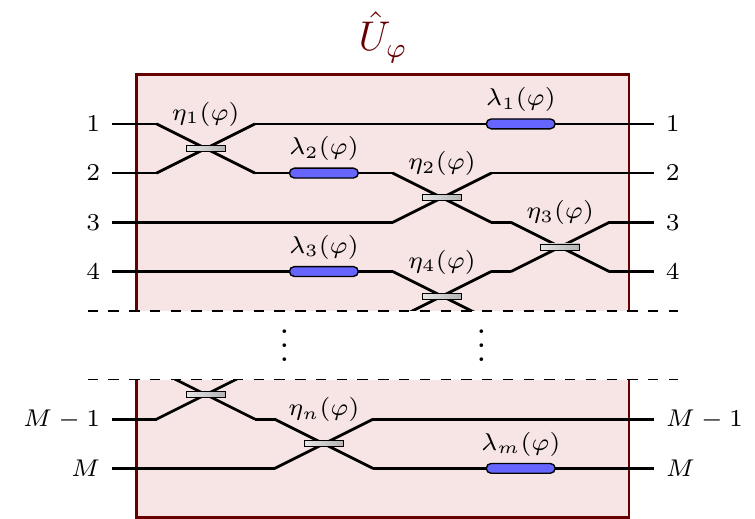}
	\caption{An $M$-mode passive linear interferometer composed of an arbitrary number of beam splitters and phase shifters having different and generic dependencies on a parameter $\varphi$.
		This scheme may represent for example the situation where $\varphi$ is the temperature or the magnitude of an electromagnetic or a gravitational field affecting the optical properties of the interferometer components.}
	\label{fig:Distributed2}
\end{figure}

The situation most commonly considered in quantum metrology is the estimation of an optical phase~\cite{Shapiro84,Giovannetti2004,monras2006,Dowling2008,Oh2019} or a phase-like parameter~\cite{Giovannetti2006,Giovannetti2011}, that, e.g., is encoded through a unitary evolution generated by a $\varphi$-independent Hermitian operator. Results thereby obtained also apply to situations in which other quantities of interest (e.g. a distance) can be converted into an optical phase~\cite{Dowling2008}, but they do not cover the estimation of physical quantities which may be spread over different components of a multi-mode interferometer (see~\figurename{~\ref{fig:Distributed2}}). 
This could be the case for the estimation of the magnitude of an external field through its influence on the optical properties of the interferometer components. For example, temperature has been used to tune the reflection and transmission coefficients of beamsplitters in on-chip interferometers~\cite{Pruessner:07,Flamini2015aa}. 
Temperature can also be used to change the optical path length through a material of index of refraction. Thus, the effects of one parameter (temperature in this case) are distributed across a network that consists of  beam splitters and phase shifters.   
Such general encodings of $\varphi$ into the interferometer give rise to parameter-dependent generators, which are scarcely considered in literature. 
Recently, some progress has been made in this direction for the estimation of a generic parameter encoded into a multi-mode passive linear interferometer~\cite{Matsubara_2019}. On the downside, it turns out that this general scenario introduces a complication in the estimation protocol which becomes highly adaptive, since \emph{both} the optimal input state of the probe, \emph{and} the optimal measurement to be performed depend on the  unknown value of $\varphi$. Therefore, this scheme, as well as previous proposals,  still relies on the use of iterative procedures, with a sequence of trial input states and measurements. This unfortunately involves an unquantified use of additional resources. This challenge is also relevant in the context of distributed quantum metrology with multiple unknown parameters and  has been overcome only if constraints in the range of variation of the parameters are given \cite{Gatto2019,Gatto2020}. Can such a serious drawback be overcome in order to ultimately implement quantum technologies for distributed quantum-enhanced metrology?

\begin{figure}[t]
	\centering
	\includegraphics{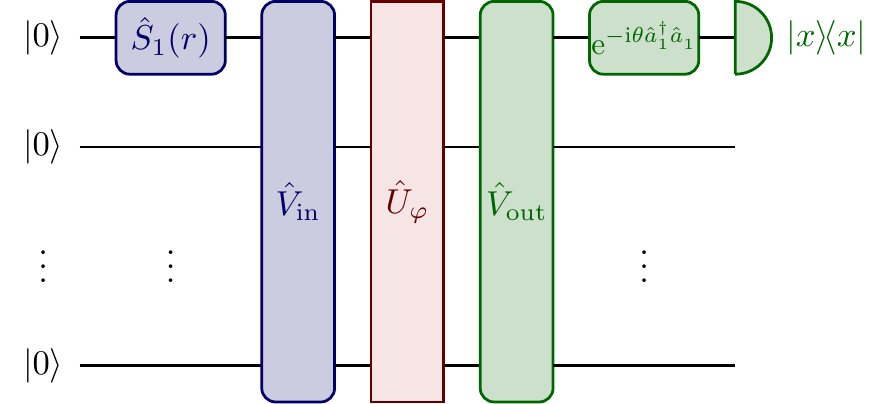}
	\caption{Block diagram of the proposed setup. The parameter $\varphi$ to be estimates is encoded in the passive linear optical interferometer $\hat{U}_\varphi$. The input state is obtained by first squeezing the vacuum in a single mode with $N$ photons in average, and then using a passive linear optical stage $\hVi$ to scatter the photons over all the interferometric channels. At the output of the interferometer, another passive linear stage $\hVo$ is employed to refocus most of the photons into the only observed channel, say the first, so that condition \eqref{eq:ucondition} is satisfied. Finally, the quadrature field $x_\theta$ is measured by means of homodyne detection, and the phase of the local oscillator shall be tuned to a value $\theta = \theta_\varphi$ which satisfies condition \eqref{eq:thetacondition}}.
	\label{fig:Generic setup}
\end{figure}

In this work, we show that this is possible by introducing a novel and experimentally feasible scheme achieving Heisenberg scaling in the estimation of a generic parameter distributed over an arbitrary passive linear optical interferometer. Independent of the structure of the multi-mode interferometer, we always pick a single mode squeezed vacuum as the only input probe and homodyne detection as measurement, while adding two passive linear optical stages before and after the interferometer, as shown in~\figurename{ \ref{fig:Generic setup}}. In our scheme, the adaptation consists in suitably choosing the unitary stages  $\hVi$ and $\hVo$ in order to achieve the Heisenberg scaling, and we show that it can be carried out only on one side of the interferometer, leaving the other stage arbitrary.
Moreover, with a careful analysis of the Fisher Information~\cite{cramer1999mathematical} associated with our scheme, performed and discussed in depth in our associated work~\cite{GrTrFaNaTa}, we show the robustness of the protocol against imperfections in $\Vi$ and $\Vo$ stages. Finally, we show that the adaptive procedure can be carried out with a knowledge of the parameter which is achievable by only classical means with $N$ probes. More precisely, we show that such a preliminary estimation affected by the standard shot-noise error $\delta\varphi = O(1/\sqrt{N})$ is sufficient to design a protocol achieving the Heisenberg scaling $\delta\varphi=O(1/N)$.
Noticeably, our estimation scheme is experimentally feasible since it employs only Gaussian states and  measurements, which are easier to manipulate and implement with respect to other states and measurements commonly used in literature.

\vspace{2mm}

\paragraph{Setup} Let us consider a given $M$-channel passive linear interferometer which depends on the parameter $\varphi$ to be estimated. The action of the interferometer on the input states is described by a unitary operator $\hat{U}_\varphi$.
The preparation of the input probe consists of two steps: first, we inject a single-mode squeezed vacuum state in the first port of a unitary stage $\hVi$, which is used to scatter the photons injected among all the modes. The input state in our protocol is therefore given by $\ket{\psi}= \hVi \hat{S}_1(r)\ket{\mathrm{vac}}$, where $\hat{S}_1(r)=\e^{\frac{r}{2}(\hat{a}_1^{2}-\hat{a}_1^{\dagger 2})}$ is the single-mode squeezing operator with squeezing parameter $r>0$, and $\ket{\mathrm{vac}}$ is the $M$-channel vacuum state. The average number of photons injected in the apparatus is thus $N=\sinh^2 r$. At the output of the interferometer, the unitary $\hVo$ is applied in order to refocus all the photons into a single mode, namely the first one, in order to capture all the information about the parameter in a single channel. In such a way the estimation can be carried out with a single homodyne detection performed on the aforementioned channel. If the refocusing procedure is not perfect there will be some probability of scattering photons into other channels, which is quantified by $1 - P_\varphi$, with 
\begin{equation}
\label{eq:P}
P_\varphi = \abs{(\Vo U_\varphi \Vi)_{11}}^2 
\end{equation} 
being the probability that a photon comes out from the first port. Hence, the number of photons into other modes different than the first one is given by $(1-P_\varphi)N$. Ideally, we would like to exploit the information encoded by the interferometric evolution in all the photons within the injected squeezed state. This corresponds to the condition $P_\varphi=1$: we are essentially channelling all the information about the parameter in a single output channel, namely the first one. Then, a homodyne detection of the field quadrature $\hat{x}_\theta$ is performed on the first channel, where $ \theta $ is the reference phase of the local oscillator employed to perform the measurement. Let 
\begin{equation}\label{eq:gamma}
\gamma_\varphi= \arg[(\Vo U_\varphi \Vi)_{11}]
\end{equation} 
be the phase accumulated through the whole setup by the field at the first output port, which will be assumed such that $\partial_\varphi \gamma_\varphi \neq 0$. The latter assumption means that the phase $\gamma_\varphi$ is not constant around the  value $\varphi$ of the parameter, which is instead effectively encoded in $\gamma_\varphi$, as a small variation of $\varphi$ implies a proportional variation of $\gamma_\varphi$. The squeezed direction of the probe at the output will be $ \gamma_\varphi \pm \pi/2$, so that the minimum uncertainty quadrature field is $ \hat{x}_{\gamma_\varphi+\pi/2} $. 
	
\vspace{2mm}	
	
\paragraph{Heisenberg scaling} The ultimate precision $\delta\varphi$ achievable in a given estimation scheme is determined by the Fisher Information $F(\varphi)$ through the Cramer-Rao Bound $\delta\varphi \geq 1/\sqrt{F(\varphi)}$~\cite{cramer1999mathematical}. Evaluating $F(\varphi)$ associated with the described setup, we find that the Heisenberg scaling can be asymptotically achieved for large $N$ if the following conditions are satisfied \cite{GrTrFaNaTa}:
\begin{align}\label{eq:thetacondition}
\theta_\varphi &\sim \gamma_\varphi \pm \frac{\pi}{2} + \frac{k_\varphi}{N},\\
\label{eq:ucondition}
P_\varphi&\sim 1-\frac{\ell_\varphi}{N},
\end{align}
where $\ell_\varphi\geqslant 0$ and $k_\varphi\neq 0$ are arbitrary but both independent of $N$, and where $\theta_\varphi$ is an optimal choice for $\theta$. From a physical point of view, $\ell_\varphi$ represents the average number of photons scattered  into channels which are not measured, while $k_\varphi/N$ represents the ``resolution'' needed in the  homodyne detection. In practice, one can even fix $k_\varphi$ to a constant value without using additional resources.

Under conditions~\eqref{eq:thetacondition} and~\eqref{eq:ucondition}, the Fisher information asymptotically reads \cite{GrTrFaNaTa}
\begin{equation}
F(\varphi)\sim 8\varrho(k_\varphi,\ell_\varphi)(\partial_\varphi \gamma_\varphi)^2 N^2
\end{equation} 
where $\varrho(k,\ell)=[8k/(16k^2 +4\ell+1)]^2$. Then, according to the Cramer-Rao bound the ultimate precision achievable with this setup is given by $\delta\varphi=O(1/N)$.  The prefactor $\varrho(k,\ell)$ reaches its maximum value $\varrho = 1$ at $k=\pm\
1/4$ and $\ell = 0$, while it vanishes at $k=0$, hence the requirement $k_\varphi\neq0$ needed to reach Heisenberg scaling.
At $k=0$ the quadrature field being measured has the minimal variance, so a vanishing Fisher Information for this value of $k$ may appear counter-intuitive. However, this occurs as a consequence of the fact that, for a local-oscillator phase $\theta =\gamma_\varphi\pm \pi/2$, the probability distribution based on a homodyne measurement in the first output channel is locally insensitive to variations of~$\varphi$. Indeed, when condition \eqref{eq:ucondition} holds, the output state in the first channel is essentially a vacuum squeezed state rotated by the phase $\gamma_\varphi$ in \eqref{eq:gamma} accumulated through the interferometer. More precisely, the probability distribution depends only on the variance of $\hat{x}_\theta$, which has a minimum for this value of $\theta$, hence being a stationary point.

The aforementioned conditions~\eqref{eq:thetacondition} and~\eqref{eq:ucondition} imply an adaptive procedure, since they depend on the true value of the unknown parameter $\varphi$. However, condition~\eqref{eq:thetacondition} only establishes the minimal resolution required in the variation of $\theta$ during the feedback procedure of the homodyne detection, and, quite interestingly, condition~\eqref{eq:ucondition} can be satisfied by manipulating only one of the two unitary stages, while leaving the other one arbitrary. The Heisenberg sensitivity is preserved even if a ratio $\ell_\varphi/N$ of the photons in the squeezed input probe is not detected in the first channel, meaning that our protocol is robust against imperfections in the adaptively optimized stage. This stage can thus be efficiently built even if the prior knowledge of $\varphi$ is affected by some error $\delta\varphi$.  

Noticeably, this uncertainty $\delta\varphi$ is allowed to be of the order $1/\sqrt{N}$ to satisfy condition~\eqref{eq:ucondition}. Hence, a classical estimation of $\varphi$ employing only $N$ photons is sufficient to gather the information needed for the adaptive optimization.
This result is due to the very structure of $P_\varphi  = \abs{(\Vo U_\varphi \Vi)_{11}}^2$, which is essentially nothing but a transition probability $P  = |\braket{v_{\mathrm{out}}}{v_{\mathrm{in}}}|^2$ between $\ket{v_{\mathrm{in}}}= U_\varphi \Vi \ket{e_1}$ and $\ket{v_{\mathrm{out}}}= \Vo^\dagger \ket{e_1}$, with $\ket{e_1}=(1,0,\dots,0)^T$. 
A simple geometrical consequence of this expression is that a small tilt of order $\delta\varphi$ between the unit vectors $\ket{v_{\mathrm{in}}}$ and $\ket{v_{\mathrm{out}}}$ yields a quadratic reduction of their transition probability $P = \cos^2 \delta\varphi \sim 1-\delta\varphi^2$. 

Furthermore, given that for any unknown parameter $\varphi$ no prior knowledge of its value is required with higher precision than $1/\sqrt{N}$, the adapted interferometer is also able to monitor the value of the parameter in an overall interval of the same order with Heisenberg-limited precision without any further change of the adapted stage.

\begin{figure}[t]
	\centering
	\includegraphics{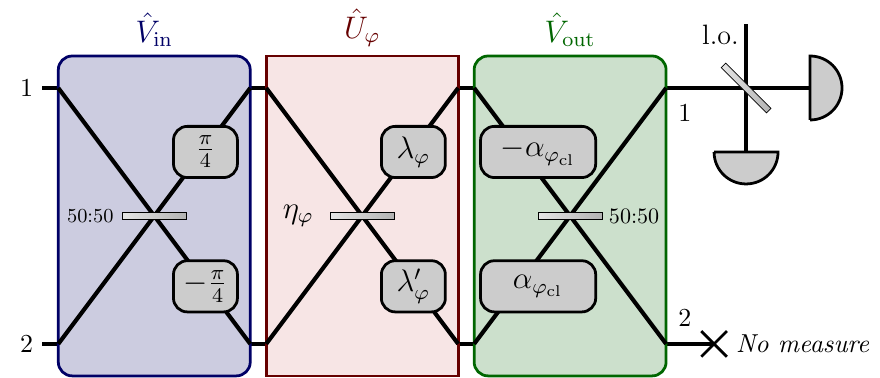}
	\caption{A two-channel example. The parameter $\varphi$ is encoded into the reflectivity $\sin\eta_\varphi$ of a beam splitter, and into the phase shifts $\lambda_\varphi$ and $\lambda_\varphi'$ associated with its two arms. A non-adapted choice of $\hVi$ is shown in the figure, realized with a beam splitter and two $\pm\frac{\pi}{4}$-phase shifts not depending on $\varphi$. The adaptation here is performed only on $\hVo$ through the tuning of ${\alpha_{\varphi_{\mathrm{cl}}}=(\lambda_{\varphi_{\mathrm{cl}}}-\lambda'_{\varphi_{\mathrm{cl}}})/2-\pi/4}$, where $\varphi_{\mathrm{cl}}$ is a prior classical estimation of $\varphi$. Eventually, homodyne detection is performed on the first output channel, with the local oscillator (l.o.) phase $\theta_\varphi$ shown in~\eqref{eq:thetaExample}.}
	\label{fig:Example}
\end{figure}

\vspace{2mm}

\paragraph{Example}
We show how our results, which are valid for an arbitrary M-port interferometer, can be applied to a particular example of a parameter $\varphi$ distributed over a $2$-channel interferometer, shown in the red box in~\figurename{ \ref{fig:Example}}. In this setup, both the reflectivity $\sin\eta_\varphi$ of a beam splitter, and the optical path lengths $\lambda_\varphi$ and $\lambda'_\varphi$ in the two arms depend on the parameter to be estimated: we can think of the parameter $\varphi$ as the magnitude of an external field, or of a characteristic of the environment, say the temperature, which in turn influences the optical properties of the devices. The functional dependence of $\eta_\varphi$, $\lambda_\varphi$ and $\lambda'_\varphi$ on $\varphi$ is assumed to be smooth. The distributed nature of $\varphi$ prevents us from thinking of it as a generalized phase, a case commonly studied in literature: hence, most of the already known results regarding quantum metrology cannot be applied. 

The unitary matrix describing  phase shifts $\lambda$ and $\lambda'$ on the two arms is the $2\times2$ diagonal matrix $U_{PS}(\lambda,\lambda') = \operatorname{diag}( \e^{\ii\lambda}, \e^{\ii\lambda'})$, while the action of a beam splitter with reflectivity $\sin\eta$ is given by $U_{BS}(\eta)=\e^{\ii \eta \sigma_y}$, with $\sigma_y$ being the second Pauli matrix. Thus, the interferometer in~\figurename{~\ref{fig:Example}} is described by  $U_\varphi=U_{PS}(\lambda_\varphi,\lambda_\varphi')\,U_{BS}(\eta_\varphi)$.

As previously discussed, Heisenberg scaling can be achieved by suitably adapting one of the two passive linear optical stages $\hVi$ and $\hVo$.
Condition~\eqref{eq:ucondition} is satisfied here with the arbitrary choice of $\hVi$ which is shown in~\figurename{ \ref{fig:Example}}. It consists of a balanced beam splitter, followed by two $\pm\frac{\pi}{4}$-phase shifts, one on each arm, and thus is described by the unitary matrix $\Vi=U_{PS}(\pi/4,-\pi/4)\,U_{BS}(\pi/4)$. The  stage $\hVo$, which will have to be adapted,  consists of two phase shifts, $\mp\alpha$, followed by another balanced beam splitter, and corresponds to the unitary matrix $\Vo=U_{BS}(\pi/4)\,U_{PS}(-\alpha,+\alpha)$. 

A direct computation of the matrix element $(\Vo U_\varphi \Vi)_{11}$ gives for this scheme the probability~\eqref{eq:P},
\begin{equation}
	P_\varphi = \dfrac{1}{2}\left(1+\sin(\lambda_\varphi-\lambda'_\varphi-2\alpha)\right),
	\label{eq:PExample}
\end{equation}
and the accumulated phase~\eqref{eq:gamma},
\begin{equation}
\gamma_\varphi = \frac{\lambda_\varphi + \lambda'_\varphi}{2} + \eta_\varphi + \frac{\pi}{2}.
\label{eq:gammaex}
\end{equation}
The adaptive procedure in this example can be accomplished by simply tuning the phase shifts $\pm\alpha$ (see \figurename{~\ref{fig:Example}}) to $\pm\alpha_\varphi$, with $\alpha_\varphi = (\lambda_\varphi - \lambda'_\varphi)/2-\pi/4$, so that 
$P_\varphi=1$. 

Of course, tuning $\alpha$ requires a prior knowledge of the parameter we want to estimate. However, as discussed above, for any arbitrary given network $U_\varphi$, by denoting with $\delta\varphi = \varphi_{\mathrm{cl}} - \varphi$ the difference between a previous coarse estimation $\varphi_{\mathrm{cl}}$ and the true value $\varphi$ of the parameter, a precision $\delta\varphi =O( 1/\sqrt{N})$ is sufficient to reach Heisenberg scaling.
Indeed, by tuning the phase shifters in the output stage according to the coarse estimation of the parameter, equation~\eqref{eq:PExample} reads $P_\varphi=[1+\cos(\lambda_{\varphi}-\lambda_{\varphi_{\mathrm{cl}}}- \lambda'_{\varphi}+\lambda'_{\varphi_{\mathrm{cl}}})]/2$. Thus, a Taylor expansion for small values of $\delta\varphi$ shows that
\begin{equation}
P_\varphi \sim 1 - \frac{1}{4}\left(\frac{\partial(\lambda_\varphi-\lambda'_\varphi)}{\partial\varphi}
\right)^2 \delta\varphi^2.
\end{equation}
It is clear from this expression that it is possible to satisfy equation~\eqref{eq:ucondition} with $\delta\varphi\sim 1/\sqrt{N}$, which is the classical precision achievable employing $N$ photons in the shot-noise limit. 

Finally, in accordance with equations~\eqref{eq:thetacondition} and~\eqref{eq:gammaex}, the phase $\theta$ of the local oscillator in the homodyne detection must then be tuned according to the value
\begin{equation}
\theta_\varphi \sim \frac{\lambda_\varphi + \lambda'_\varphi}{2} + \eta_\varphi + \frac{k_\varphi}{N}.
\label{eq:thetaExample}
\end{equation}
We  notice that, although not appearing in $\hVo$, the value of the unknown reflectivity $\sin\eta_\varphi$ influences the quadrature field to be measured.

\vspace{2mm}

\paragraph{Conclusions} We provided an experimentally feasible scheme for the estimation of a generic parameter encoded into a $M$-mode passive linear interferometer with Heisenberg scaling precision. Our proposal could find applications in those situations where the parameter is distributed among different components of the interferometer, and as an example it can provide a novel paradigm in quantum thermometry~\cite{PRLQuantTherm} or quantum magnetometry~\cite{PRAQuantMag,bhattacharjee2020quantum}, in those situations where an external field such as the temperature or a magnetic field influences the optical properties of beam splitters and phase shifts in an interferometer. The practical advantages of our scheme are twofold: on one hand, it employs only a Gaussian state and Gaussian measurements, whose experimental feasibility is widely known; on the other hand, the adaptive procedure is facilitated by reducing the adaptivity to only one stage (either in the input probe or in the measurement) of the estimation scheme. Moreover, the amount of information on the parameter required to perform this adaptation is achievable with any classical shot-noise limited estimation. Finally, no further parameter-dependent adaptation is necessary to measure with Heisenberg-limited sensitivity any value of the parameter within an overall range of variation of the order of $1/\sqrt{N}$. To the best of our knowledge, our estimation protocol for a distributed parameter is the first one requiring a feasible input states for the probe, and such a weak adaptation.

\section*{Acknowledgements}
This work was supported by the Office of
Naval Research Global (N62909-18-1-2153). PF and GG are partially supported by Istituto Nazionale di Fisica Nucleare (INFN) through the project “QUANTUM”, and by the Italian National Group of Mathematical Physics (GNFM-INdAM).

GG and DT contributed equally to the drafting of this work.

\nocite{*}
\bibliographystyle{apsrev4-1} 
\bibliography{references}

\end{document}